\documentclass[12pt]{article}

\usepackage{graphicx}
\usepackage{wrapfig}
\graphicspath{ {./images/} }	
\usepackage[vmargin=1in,hmargin=1in]{geometry}
\usepackage{hyperref}
\usepackage{caption}
\usepackage{subcaption}
\usepackage{enumitem}
\usepackage{fancyhdr}

\makeatletter
\renewcommand{\maketitle}{\bgroup\setlength{\parindent}{0pt}
	\begin{flushleft}
		\textbf{\@title}
		
		\@author
	\end{flushleft}\egroup
}
\makeatother

\title{\textbf{{\Large Astro2020 Activity, Project of State of the Profession Consideration (APC) White Paper: \newline \newline Title: All-Sky Near Infrared Space Astrometry \newline \newline State of the Profession Considerations: Development of Scanning NIR Detectors for Astronomy}}}
\date{\today}
\author{\textsc{\\}
		\textsc{\bf Authors:\\}
		\textsc{\\}
		\textsc{\small David Hobbs, Lund Observatory, Box 43, SE-221 00, Lund, Sweden}\\
		\textsc{\small Christopher Leitz, MIT Lincoln Laboratory, 244 Wood St., Lexington, MA 02421, USA}\\
		\textsc{\small Jo Bartlett, Mullard Space Science Lab, UCL, Holmbury St. Mary, Dorking, Surrey, RH5 6NT, UK}\\
        \textsc{\small Ian Hepburn, Mullard Space Science Lab, UCL, Holmbury St. Mary, Dorking, Surrey, RH5 6NT, UK}\\
		\textsc{\small Daisuke Kawata, Mullard Space Science Lab, UCL, Holmbury St. Mary, Dorking, Surrey, RH5 6NT, UK}\\
		\textsc{\small Mark Cropper, Mullard Space Science Lab, UCL, Holmbury St. Mary, Dorking, Surrey, RH5 6NT, UK}\\
		\textsc{\small Ben Mazin, Department of Physics, University of California, Santa Barbara, CA 93106, USA}\\
		\textsc{\small Anthony Brown, Leiden Observatory, Leiden University, Niels Bohrweg 2, 2333 CA Leiden, The Netherlands}\\
		\textsc{\small Valeri Makarov, US Naval Observatory, 3450 Massachusetts Ave NW, Washington DC 20392-5420, USA}\\   
		\textsc{\small Barbara McArthur, McDonald Observatory, University of Texas at Austin, Austin, TX 78712-1205, USA}\\ 
	    \textsc{\small Anna Moore, The Australian National University, Canberra, Australia}\\	
	    \textsc{\small Robert Sharp, The Australian National University, Canberra, Australia}\\
	    \textsc{\small James Gilbert, The Australian National University, Canberra, Australia}\\
	    \textsc{\small Erik H{\o}g, Niels Bohr Institute, Blegdamsvej 17, 2100 Copenhagen {\O}, Denmark} \\
}

\fancypagestyle{style1}{
	\fancyhf{}
	
	\fancyfoot[L]{}
}

\fancypagestyle{style2}{
	\fancyhf{}
	\fancyfoot[L]{\tiny\it DISTRIBUTION STATEMENT A. Approved for public release. Distribution is unlimited. \\
			This material is based upon work supported by the Under Secretary of Defense for Research and Engineering under Air Force Contract No. FA8702-15-D-0001. Any opinions, findings, conclusions or recommendations expressed in this material are those of the author(s) and do not necessarily reflect the views of the Under Secretary of Defense for Research and Engineering.}
	
}

\begin{document}
\maketitle

\vspace{0.3in}
\noindent
{\bf Contact information for lead author:}\\
\\
David Hobbs \\ Lund Observatory \\ Box 43, SE-221 00 \\ Lund \\ Sweden \\ Email: david@astro.lu.se \\ Tel.: +46-46-22\,21573

\newpage

\pagestyle{style1}

\section{Introduction}\label{introduction}
Gaia is a revolutionary space mission developed by ESA and is delivering 5 parameter astrometry, photometry and radial velocities over the whole sky with astrometric accuracies down to a few tens of micro-arcseconds. A weakness of Gaia is that it only operates at optical wavelengths. However, much of the Galactic centre and the spiral arm regions, important for certain studies, are obscured by interstellar extinction and this makes it difficult for Gaia to deeply probe. This problem can be overcome by switching to the Near Infra-Red (NIR) but this is not possible with silicon CCDs. Additionally, to scan the entire sky and make global absolute parallax measurements the spacecraft must have a constant rotation and this requires the detectors operate in Time Delayed Integration (TDI) mode or similar.

A clear improvement on Gaia is to go into the NIR and such a proposal has been made to ESA (Hobbs, et al., 2016, \cite{2016arXiv160907325H}) and ESA subsequently studied the mission concept in detail (\href{http://sci.esa.int/future-missions-department/60028-cdf-study-report-gaianir/}{GaiaNIR – Study to enlarge the achievements of Gaia with NIR survey}). The ESA study found that a TDI solution would give similar accuracies as Gaia despite doubling the wavelength range. However such TDI technology for NIR detectors scarcely exists mainly because it is not possible to transfer electrons from pixel to pixel in typical materials used for NIR detectors. The science cases for GaiaNIR have been outlined by Hobbs, et al., 2016, \cite{2016arXiv160907325H} and in a proposed collaboration on this project with the US recently submitted as an Astro2020 Science white paper on ``All-Sky Near Infrared Space Astrometry", McArthur, et al., 2019, \cite{2019arXiv190408836M}. Additionally a further white paper on this topic will be submitted to ESA’s continuation call of Cosmic Vision, called \href{https://www.cosmos.esa.int/web/voyage-2050}{Voyage 2050}, in August of this year. The science return from such a mission is very promising but a solution to the technology problem of implementing a TDI like solution in large format NIR detectors must now be found.


\section{Possible solutions}\label{solutions}
Astronomical-grade infrared detectors are well established for both ground- and space-based applications. At shorter IR wavelengths, the premium detectors are HgCdTe devices, especially those fabricated by Teledyne in the USA. Unlike CCD detectors, the state-of-the-art for optical wavelengths, the HgCdTe devices cannot be used in a scanning mode, in which the image is moved within the device synchronously with the movement of the scanning of the optical system (i.e. TDI). TDI is useful in many applications, and particularly for surveys, as individual exposures are not required, and data simply pour out of the detector array in a continuous stream. An example of TDI in ground-based astronomical applications is the Sloan Digital Sky Survey (SDSS), one of the most influential astronomy initiatives ever while a space-based example is ESA’s Gaia mission mentioned above. Such surveys could not have been achieved without TDI operation but have been limited to the optical band.

This is a significant limitation. The importance of NIR in modern astronomy and cosmology is abundantly clear with most major ground-based facilities operating powerful infrared instrument suites, and major upcoming missions such as James Webb Space Telescope, Euclid and WFIRST are either completely or significantly orientated to infrared observations. There is a compelling case, therefore, for an infrared detector that can operate in TDI.

There are a number of possible approaches to developing TDI-NIR detectors:
\begin{enumerate}[itemsep=1pt,parsep=0pt]
\item A hybrid solution which uses a HgCdTe NIR detector layer bump bonded to a Si CCD. The idea is that the photons are detected in the surface NIR layer and transferred to the Si buried channel at each pixel. Charge can then be easily moved 
along the pixels of the same column in sync with the charge generation, thus achieving TDI. What is not known yet is how efficiently the charge can be transferred from the NIR detection layer to the Si CCD and if both materials can be operated at the same temperature. 
\item Using HgCdTe Avalanche Photodiodes (APDs) with TDI-like signal processing capability. The challenge here is to scale the technology to large format arrays and ensure the dark current does not introduce unwanted noise at temperatures above 100 K.  
\item Ge detectors due to the lower band gap can detect NIR radiation of longer wavelengths than possible with Si detectors \cite{LeitzGe2018}. Clearly this technology is new but many of the manufacturing techniques developed for Si are also applicable to Ge and further development is needed to see if they can be used for our application with low noise and visible-NIR capabilities in large format arrays.
\item Microwave Kinetic Inductance Detectors (MKIDs) are cooled, multispectral, single photon counting, multiplexed devices, capable of observation in the UV 
through to SWIR 
They measure the energy of each photon to within several percent and log the time of arrival to within 2 microseconds, making them ideal for TDI like operation. Whilst relatively new, small MKID arrays have already been utilised on ground-based telescopes for astronomy, demonstrating their potential.
\end{enumerate}
The purpose of this white paper is to highlight possible technology for a NIR Gaia-like astrometry mission. A number of approaches have already been identified but detailed studies are needed in the coming years to assess the best approach and to develop laboratory demonstrators before full scale production of such devices can be undertaken.

\section{HgCdTe NIR detector bump bonded to a Si CCD}\label{HyCCD}

\subsection{Introduction}

\begin{wrapfigure}{r}{0.45\textwidth}
	\centering
	\includegraphics[width=0.45\textwidth]{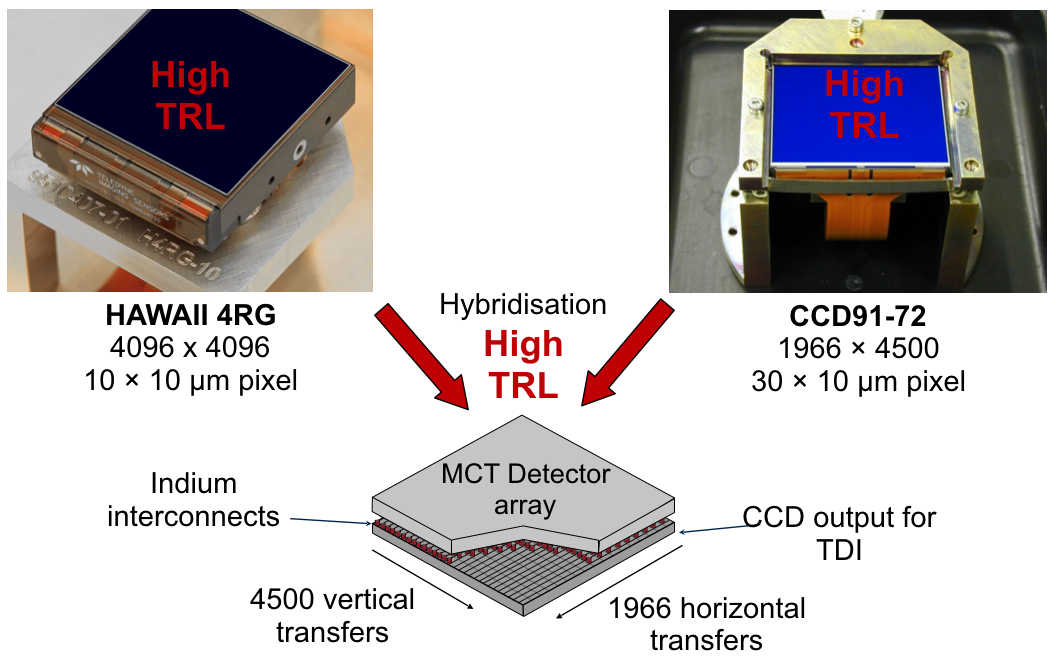}
	\caption{Overall concept for the methodology to produce a HyTED device. Hybrid system of MCT and CCD would allow charge domain TDI in NIR.\label{fig:HyTED1}}
\end{wrapfigure}

The recent acquisition by Teledyne of e2v brings together the world’s premium infrared and optical detector manufacturers. Whatever the advantages or disadvantages this may present to the astronomy community, one outcome is that Teledyne-e2v is seeking to merge their intellectual property to maximise their mutual benefit. Hence, a new hybrid detector which combines the infrared detection capability of their HgCdTe devices with the TDI capability of transferring single electrons in their CCD devices is of interest, and GaiaNIR provides an opportunity to explore whether such a hybrid NIR TDI detector may be feasible. They have adopted the name, Hybrid Teledyne-e2v Detector (HyTED). 

\subsection{Methodology}

Mercury Cadmium Telluride (MCT) detectors use a HgCdTe detector layer bump-bonded to silicon CMOS circuitry using indium beads. The proposal is to replace the CMOS circuitry with the highly developed CCD structure, so that the indium bonds interface directly with the CCD pixels, to transfer electrons or holes directly into them, after which they can be clocked to the readout register as in a normal CCD in TDI operation (see Fig. \ref{fig:HyTED1}).

To bump bond a HgCdTe detector plate to a Si CCD structure conducting Indium beads are normally used. The problem is that transferring one electron (or hole) into the top side of the bead gives some statistical number of electrons (or holes) out of the bottom. If you want one electron (hole) at the top to produce one electron (hole) at the bottom, the bead has to be a semiconductor biased in the correct manner. This statistical broadening of the signal is an extra noise term and is normally fine if there are many electrons, however with TDI operation, electrons (holes) are transferred one at a time, and this noise broadening caused by the Indium being a conductor engulfs the single electron (hole). The critical technology to develop is to find a semiconductor replacement for the Indium beads which is then biased in such a way that an electron (hole) in at the top produces one out at the bottom. 

\begin{wrapfigure}{r}{0.45\textwidth}
	\centering
	\includegraphics[width=0.45\textwidth]{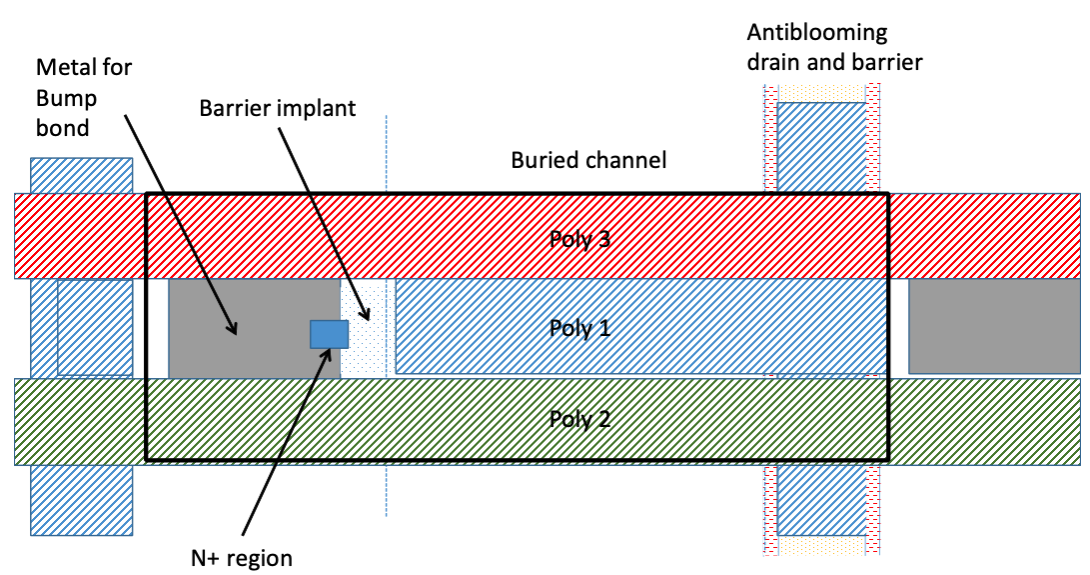}
	\caption{Proposed pixel structure. A pixel is enclosed in the black rectangle. Poly is an abbreviation of polysilicate, which is the material from which the electrodes are constructed. The figure shows the metal contact for the bump bond, and the 3 CCD electrodes (poly 1, 2 and 3) used to transfer charge through the CCD.\label{fig:HyTED2}}
\end{wrapfigure}

The proposed CCD pixel structure to accept the charge from the HgCdTe layer via the indium bead is shown in Fig. \ref{fig:HyTED2}. This is a 3-phase device. There is a metal pad between electrode 2 and 3 to contact the indium bead. While the structure can be designed with standard techniques to provide the pixel sizes, quantum efficiencies, noise levels, readout rates, dark currents and auxiliary structures such as supplementary buried channels and anti-blooming drains for astronomical use, the transfer of single photo-electrons from the HgCdTe layer via the indium bead is not yet feasible. No such transfer occurs in the current Teledyne devices, so the problem doesn't arise: instead, the photoelectrons remain in the HgCdTe layer, to be discharged at the end of the exposure by the CMOS circuitry associated with each pixel on the other side of the indium bead, and the measured flux is then proportional to the discharge. A second consideration is that the transfer is then of electrons upwards, or holes downwards, rather than electrons, so that the CCD layer needs to be of p-channel Si, rather than the n-channel devices mostly used in astronomical application. This is, however, less problematic, as e2v have had a long history of developing CCDs using p-channel technology which have excellent performance and generally superior tolerance to radiation damage.

The main challenge is the transfer of information at the single electron level into the CCD. In TDI mode, this is particularly severe: as the image scans over the detector, counts above the background level fall first into one pixel, then the next and the next, so the integration time on each pixel may be as short as a millisec, with correspondingly low charges to be transferred – occasionally single transfers. The accumulation of charge occurs at the CCD pixel level, and not in the HgCdTe layer. Related is the time constant for the transfer: if this is not much shorter than the pixel clocking, then the charge may enter into pixels later in the scan than those intended, creating a trail behind the image. This is associated with the persistence effects which are present in the HgCdTe devices. Other challenges relate to the optimisation of the operating temperature, as lower temperatures are favoured to minimise the dark current produced in the HgCdTe layer, while warmer temperatures are better for maximising the charge transfer in the Si in the presence of radiation damage. The envisaged development programme would be as follows:
\begin{enumerate}[itemsep=1pt,parsep=0pt]
\item Investigate the physical characteristics of barriers to the transfer of charge from HgCdTe layer to a Si pixel. Is this caused by electropotentials at interfaces to and from the indium beads, or because the indium conductor cannot be fully depleted, or some other effect; are there fundamental limits to the rates at which the holes can diffuse through the HgCdTe layer? This will require an understanding of the materials, processes and structures and close cooperation with Teledyne-e2v is needed to make progress.

\item Explore steps to enhance the electron transfer -- different materials for the pads on the CCD pixel interfacing to the indium beads (see Fig. \ref{fig:HyTED2}); replacing indium with a different material which can be fully depleted; active biasing of the HgCdTe layer with respect to the Si; retaining the current CMOS layer in the HgCdTe devices and interfacing from there into the CCD, etc.

\item Optimizing the temperature for the hybrid device. This would include the characterisation, in the presence of radiation damage, of a p-channel e2v CCD at temperatures of 100 – 140K, colder than currently utilised. The implications for some associated processes, such as glue bonds for flexi-connectors, will also need to be assessed.

\item The specification of an interim hybrid device, using a Si photodiode detection layer and n-channel CCD working at optical wavelengths, and carrying out a programme testing the performance of the transfer of electrons through the bump bonds into the pixels. A minimum of 2 devices would be desirable, to measure device-device variability.

\item The prototyping of one or more HyTED device types, and then carrying out test programmes with the aim of characterising the end-end transfer explored in (1) and (2) above. Again, the number of devices to be characterised will be dependent on the success of the processing, and on processing capacity within Teledyne-e2v, but a minimum of 2 devices would be desirable, in order to measure device-device variability.
\end{enumerate}

\section{Electron Initiated Avalanche Photodiodes (e-APDs)}\label{APDs}

APDs, semiconductor electronic devices which exploit the photoelectric effect and can be considered
the semiconductor analogue of the photomultiplier and with sub-electron readout noise, is very
promising technology with the limitation of increasing the dark current at temperatures above 100K.
The very fast read out of these devices makes APDs inherently suited to a TDI like signal processing
mode. APDs are generally fabricated using $n^+/n^-(\nu)/p^+$ structures as shown in Fig. 1 from
\cite{DOI:10.14429}. Electrons and holes in the depletion region of an APD get accelerated under the
influence of a  high electric field and gradually acquire sufficient kinetic energy to give nearly
exclusive impact ionization of the electrons, hence the name e-APDs \cite{Feautrier:15}. These
newly generated carriers also obtain enough energy from the field in reverse bias and create another
electron-hole pairs. This process is  known as avalanche multiplication. The electric field across
the depletion region separates the photo-generated electron-hole pairs and thus contributes to the
photo current. The $n^-$ region starts acting as a multiplication zone and the surrounding $p^+$
region is known as the absorption layer. Internal photocurrent gain takes place through impact
ionisation. The multiplication region of the APD plays a crucial role to achieve high avalanche gain
at low bias with low noise and high bandwidth. In general, APDs operate below the break down
voltage of the semiconductor and are known as linear amplifiers for the input optical signal,
whereas Geiger mode APDs work at greater than the break down voltage \cite{DOI:10.14429}. 

\begin{wrapfigure}{r}{0.4\textwidth}
	\centering
	\includegraphics[width=0.4\textwidth]{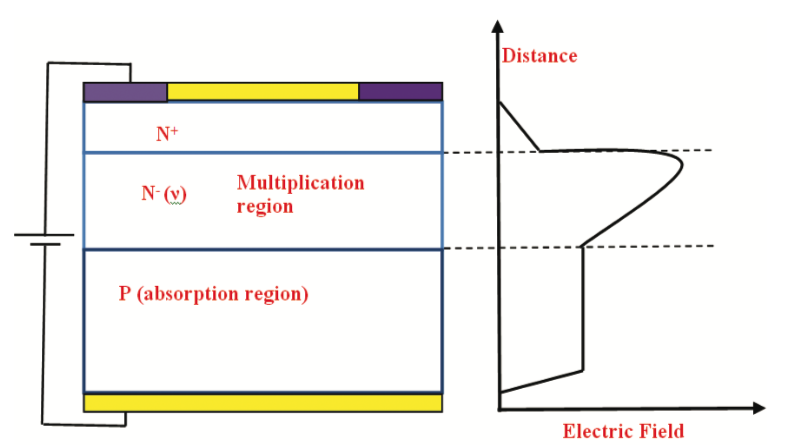}
	\caption{Fig. 1 from \cite{DOI:10.14429} shows a schematic diagram of an APD. 
		\label{fig:APD1}}
\end{wrapfigure}

The development of HgCdTe e-APDs was a significant advance in detector technology for the short-wave infrared (SWIR) to mid-wave infrared (MWIR) wavelength region \cite{Sun:17}. HgCdTe photodiodes have high quantum efficiency and low dark current at low temperature. The optimal operating temperature depends on the band-gap or the cut-off wavelength of the HgCdTe material. Recently, the photoelectron multiplication gain has reached 500 or higher, more than sufficient to overcome the electronic noise of the ROIC, while still maintaining low dark current and near unity excess noise factor. HgCdTe e-APDs operate below the breakdown voltage and provide linear analog output. They have a wide dynamic range and no dead-time and no afterpulsing. 

The European Southern Observatory (ESO) first started to explored e-APD arrays in 2008 for astronomy \cite{Leonardo:16} and they clearly showed promise but were not suitable for long integration times and low dark current. Arrays were manufactured but needed cooling to 40K and had gains limited to $\times$20 and integration periods limited to 10s of milliseconds. However in 2012 parallel developments in Metal Organic Vapour Phase Epitaxy (MOVPE) for thermal imaging had progressed to the point that prototype e-APD devices could be trialled. The first MOVPE batches showed the benefit of bandgap engineering immediately with operation at temperatures up to 100K, gains up to $\times$60 and few defects in integration times of seconds \cite{Leonardo:16}. One of the reasons e-APD arrays have matured so quickly is that these conditions are easily met using near-standard manufacturing processes. e-APDs offers voltage controlled gain at the point of photon absorption, electron gain values up to 1000, virtually zero power consumption, bandwidths to GHz, high stability, high uniformity, no impact on the pixel design and non-destructive readout schemes with subpixel sampling are possible with negligible added noise. Noise due to dark current is not included here and can be significant particularly as it is gain-amplified in many devices, so some caution with dark current is needed and it may be a limiting effect.

The focus of e-APD development for astronomical use in recent years has been on wavefront sensing.
Achieving the demanding tolerances on wavefront error control necessary to reach the diffraction
limit with the next generation of Extremely Large Telescopes (ELT) has driven the development of the
SAPHIRA e-APD \cite{Atkinson:2014, Atkinson:2016, 2018AJ....155..220A, Atkinson:2018,
2014SPIE.9148E..17F, 2016SPIE.9909E..12F, 2016SPIE.9915E..0OH}. As such the SAPHIRA will be the
detector of choice for many ELT AO wavefront sensor systems (e.g., the Giant Magellan Telescope
Integral Field spectrograph, GMTIFS, \cite{2018SPIE10702E..1VS}). Largely with this goal in mind, a
commercial product offering a turn-key solution to such imaging is available
\cite{2017SPIE10209E..0GF}, the C-RED 1 camera from
FirstLight\footnote{https://www.first-light-imaging.com/product/c-red-one/} imaging. While highly
successful \cite{96593b97fd7e4d50bb6c941d1f28c48d}, the commercial package is designed to deliver a
reliable ground-based solution but is not optimised for the weight, power and environment
constraints relevant to a space-based observatory.

On-sky performance has been demonstrated in imaging mode a number of times, notably for the purposes of “lucky imaging” \cite{Atkinson:2018, 2018SPIE10709E..23V}.  Indeed, early success with one such system at the Australian National University (ANU) has led to a space-based TDI mission for astronomy. The “Emu” mission will demonstrate space-flight readiness for SAPHIRA with a $\sim$100 mm telescope deployed on the International Space Station (ISS) which mitigates many of technical hurdles associated with deploying small payloads, instead focusing on technology demonstration. A prototype system was successfully demonstrated on-sky in April 2019.
While the current generation SAPHIRA e-APD has only a modest scale (320x256 pixels, but with a high pixel operability, approaching 100\%), a number of active collaborations are underway \cite{2016SPIE.9915E..0OH} to deliver large format devices more relevant for the extended focal plane mosaics typically needed for large survey missions.  


In conclusion, e-APDs because of their very fast readout can allow the photon counts to be signal processed to give a TDI like operation mode suitable for a scanning space mission, e.g. reference \cite{El-Desouki:14} shows TDI implemented in standard CMOS technology. 
MOVPE growth provides full control over the bandgap and doping profile with arsenic used as the acceptor and iodine used as the donor. For e-APDs it allows the absorber, p-n junction region and multiplication region to be independently optimised \cite{Leonardo:16}. The main challenges are, firstly, to extend the response of the material down to 400nm by removing a part of the detector substrate allowing visible light to also be detected while maintaining a reasonable QE. Secondly, to greatly increase the detector size format which is possible as a future technology development. There are many applications of such detectors in astronomy, Earth observation and, of course, military. For example, the Pulsed All-sky Near-infrared Optical SETI (PANOSETI) observatory will feature an array of Geiger-mode e-APDs in a wide-field NIR (950--1650 nm) instrument capable of scanning the northern hemisphere \cite{Li:19}\cite{2018SPIE10702E..5IW}.

\pagestyle{style2}
\section{Ge detectors}\label{Germanium}

\begin{wrapfigure}{r}{0.5\textwidth}
	\centering
	\includegraphics[width=0.5\textwidth]{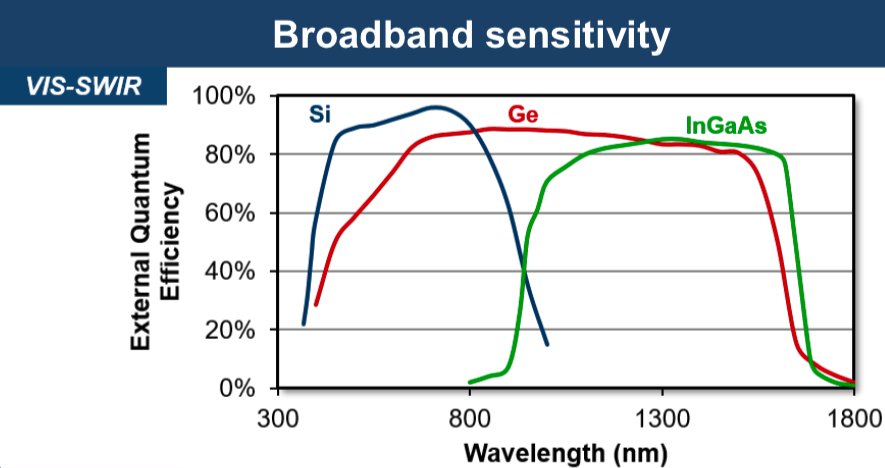}
	\caption{Comparison of the broadband response of Si, Ge and InGaAs.\label{fig:Ge1}}
\end{wrapfigure}

Silicon CCDs have found many applications in astronomy and in particular for the Gaia mission by exploiting the numerous advantages of silicon CCDs such as noiseless binning (grouping signals from adjacent pixels for enhanced sensitivity) and the ability to perform TDI to track moving objects with no noise penalty which is so useful for detecting very faint signals. However, silicon has a limited wavelength range and even red enhanced silicon detectors struggle to reach 1100 nm. Clearly any material with similar electrical properties to Si but a broader wavelength range would be ideal for GaiaNIR. Germanium is such a material, being sensitive to NIR and soft x-rays,  but is a less well developed material for use in detectors. Fig. \ref{fig:Ge1} shows a comparison of the spectral sensitivity of Si, Ge and InGaAs. InGaAs is another material which is sensitive in the NIR but (due to absorption in the InP substrate used to grow the InGaAs layers) lacks a strong overlap with the optical band while Ge is sensitive in both optical and NIR. This is ideal for our proposed mission as we also want to  re-measure the same stars as seen by the optical Gaia mission while also detecting new objects in the NIR .

Leitz, et al. \cite{LeitzGe2018} at MIT Lincoln Labs have pioneered the development of Ge CCD technology. They have shown that germanium offers higher energy resolution than silicon and because of the high mobility of both electrons and holes in germanium, output amplifiers built on germanium are expected to exhibit lower white noise than those built on silicon. Additionally, because defect-free 200-mm-diameter germanium wafers are commercially available 
(\href{https://eom.umicore.com/en/en/germanium-solutions/products/}{see Umicore Substrates, Olen Belgium}), germanium CCDs can provide these advantages while retaining compatibility with the Very Large Scale Integration (VLSI) tool set used to fabricate silicon detectors. Because this material can leverage the advanced semiconductor process tools used for silicon CCDs, and because the CCD is inherently a monolithic detector that does not require hybridization to a readout circuit, germanium CCDs matching the large formats of silicon CCDs should be attainable. 

\begin{wrapfigure}{r}{0.5\textwidth}
	\centering
	\includegraphics[width=0.5\textwidth]{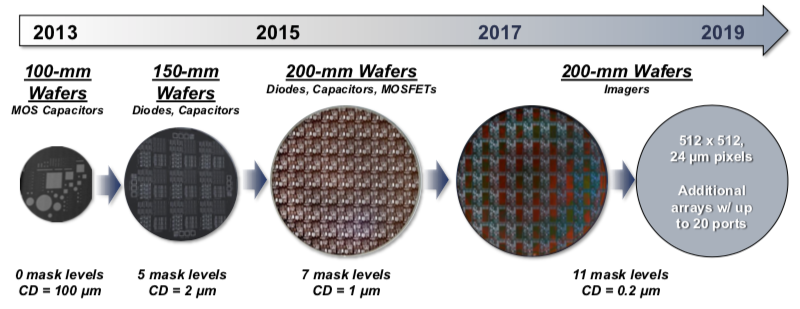}
	\caption{MIT Lincoln Laboratory has been developing germanium imagers for several years for X-ray and NIR-band imaging.\label{fig:Ge2}}
\end{wrapfigure}


A germanium CCD therefore potentially offers the unique combination of broadband response, low noise, and large detector format. A front-illuminated germanium CCD was first demonstrated in the 1970s \cite{1974ApPhL..25..747S}, and later further explored in the 1990s, but these devices suffered from a variety of limitations such as high dark-current and poor charge-transfer efficiency (CTE). In contrast, recent improvements to germanium material quality now enable germanium CCDs with performance attributes that match silicon CCDs. In particular, recent work on the growth of high-quality GeO$_2$ gate dielectrics on Ge has enabled high-quality Ge-based metal-oxide-semiconductor (MOS) capacitors, p-n junction diodes, and buried-channel MOS field-effect transistors (MOSFETs), collectively the building blocks of CCDs \cite{Leitz_2017}.


The primary challenge in fabricating low-noise CCDs on germanium is the realization of low semiconductor-dielectric interface state density required for both low dark current and high charge-transfer efficiency (CTE). Recently, mid-gap surface state density values below $10^{11}$ eV$^{-1}$ cm$^{-2}$, approaching the interfacial quality of silicon devices, have been reported for a gate dielectric stack utilizing GeO$_2$ at the insulator/semiconductor interface. These GeO$_2$ layers can be formed by thermal or plasma oxidation, and are usually capped with high relative-permittivity dielectrics to prevent subsequent degradation \cite{Leitz_2017}. 
Future Directions for Germanium CCDs focus on three areas as detectors scale from today’s small arrays to large-format devices suitable for new science missions.

\begin{enumerate}[itemsep=1pt,parsep=0pt]
\item Increase the charge-transfer efficiency. MIT Lincoln Labs have demonstrated that charge-transfer efficiency (CTE) is limited by bulk trapping, likely from trace metallic contaminants. In the absence of bulk trapping, CTE expected to be at least 99.99\%. Known methods for eliminating metallic contamination in CCDs are being used.
\item Increase the array size to Mpixel class detectors, primarily by reducing the density of gate-to-gate shorts. While the yield for today’s germanium CCDs is low, silicon CCDs fabricated in a similar but more mature process show much higher yield; an increased emphasis on using in-line metrology and inspection should close that gap.
\item Perform backside illumination for high sensitivity. 
a) All germanium CCDs to date are frontside-illuminated with metal coverage shading the front surface.
b) Wafer-scale backside illumination of silicon CCDs is commonplace, but with limited quantities of germanium wafers available MIT Lincoln Labs are first pursuing chip-level thinning.
c) MIT Lincoln Labs have developed a process to backside thin and passivate hybridized Ge diode arrays which they are currently testing on germanium CCDs.
\end{enumerate}

\section{Microwave Kinetic Inductance Detector (MKID)}\label{MKID}

Microwave Kinetic Inductance Detectors (MKIDs) are photon-counting, energy-resolving cryogenic detectors that enable simultaneous observation per pixel of photons over a wide wavelength range; they are sensitive to wavelengths from 0.2 to 5 microns (UV - mid-IR), although it is not possible to image over the entire wavelength range with a single array. MKIDs measure the individual photon energy to within several percent or better, providing multi-spectral imaging without the need for filter wheels. In addition, they measure the photon time of arrival to within 2 microseconds and unlike conventional detectors that have set exposure and readout times, MKIDs effectively video a scene.

\begin{wrapfigure}{r}{0.25\textwidth}
	\centering
	\includegraphics[width=0.25\textwidth]{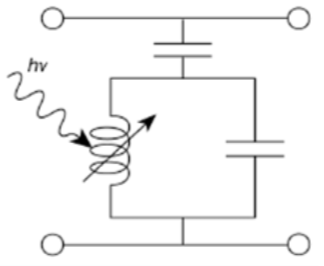}
	\caption{Electrical equivalent circuit for an MKID pixel\label{fig:MKID1}}
\end{wrapfigure}

MKIDs belong to a group of detectors known as Low Temperature Detectors (LTDs). These detectors typically operate at sub-Kelvin temperatures, often requiring 50-100 mK. There is a long history spanning $\sim$ 30 years of LTD research, covering devices such as microcalorimeters, bolometers, transition edge sensors (TES) and superconducting tunnel junctions (STJ). Despite their cooling requirements, LTDs have already been used in space on Planck (52 $\times$ bolometers, at 100 mK) and Astro-H/Hitomi (36 pixel TES at 50 mK) and will be flown on the future Athena mission (3168 pixel TES at 50 mK). In comparison, MKIDs are relatively new (10--15 years), however their development, pioneered by B. Mazin of the University of California, Santa Barbara has been rapidly progressing \cite{doi:10.1063/1.3292300} \cite{2017OExpr..2525894S}; MKID arrays have already been demonstrated by Mazin’s group on the 200" Hale Telescope at the Palomar Observatory and the Lick Observatory 3-m telescope, and now have a permanently installed instrument at the Subaru 8-m telescope. MKIDs are inherently easy to multiplex, giving them a significant advantage over other LTDs, which require either a single amplifier chain per pixel or a complex multiplexing technique. This allows substantially larger MKID arrays to be achieved; the largest array to date is 20\,440 pixels, with megapixel arrays anticipated in the near future.

\subsection{MKID fabrication and photon detection process}


MKIDs are lithographically produced, thin film superconducting detectors. Each MKID pixel is comprised of a superconducting inductor and a capacitor to form a resonant (LC) circuit, with each pixel having a different resonant frequency. Fig. \ref{fig:MKID1} shows the equivalent electrical circuit.
MKIDs utilise the kinetic inductance effect whereby incident photons change the surface impedance of the superconductor by breaking Cooper pairs into quasi-particles. A photon is detected by monitoring a constant microwave probe signal tuned to the pixel resonant frequency -- the resulting change in surface impedance translates to a change in the probe signal phase and amplitude that is proportional to the photon energy. 

Multiplexing is easily achieved by using a comb of probe signals (one frequency for each pixel) within one feedline. Using this method, 2,000 pixels can currently be read out using one feedline pair (2 coaxial cables); one feedline carries the microwave signals to the array, whilst the other carries the return signals to the read-out electronics. Therefore to read out a 20,000 pixel array, only 10 feedline pairs (20 coaxial cables) are required. This multiplexing technique gives MKIDs a significant advantage over other LTDs.

\pagestyle{style1}


\subsection{Detector cooling, readout noise and TDI}
With the exception of the detector array and the low temperature amplifier (called the High Electron Mobility Transistor, HEMT), all of the read-out electronics are at room-temperature; the array needs to be operated at 100 mK, whilst the HEMT needs to be cooled to $\sim$ 4 K. The Mullard Space Science Laboratory (MSSL) has recently demonstrated a new continuous sub-Kelvin cryo-cooler \cite{2012SPIE.8452E..1OB} that could enable continuous 100 mK cooling for a 400,000 pixel array (200 feedline pairs). There is no read-out noise associated  with the detector array itself, only the associated Poisson uncertainty in the conversion of the photon energy into quasi-particles. In terms of the readout, an incident photon alters the microwave probe signal and the corresponding signal phase and amplitude change along with the event time, is logged. The system floor noise  of the read-out electronics is set by the HEMT and the room-temperature ADC (Analogue-to-Digital converter). 

Each detected photon produces a data packet containing information regarding the photon energy and the time of arrival (to within 2 $\mu$s). MKIDs effectively video a scene as they are continuously read out -- there is no set exposure and read-out times. This makes MKIDs inherently suited to TDI like operation, enabling improved image reconstruction and minimal blurring, providing the spacecraft pointing is accurately known. MKID's energy resolution means that they can act as spectrographs opening intriguing new possibilities. Further details can be found in a dedicated Astro2020 white paper on MKIDS (MKIDs in the 2020s, Mazin, et al. 2019 \cite{Mazin2019}).
 


\subsection{Development Requirements}
The following near term developments are needed in order for an MKID NIR TDI focal plane to be realised for a possible future mission. These developments would also be required to provide an input into, and investigate, a system level design.
\begin{itemize}[itemsep=0.1mm, parsep=0pt]
\item Production of large MKID arrays:
Whilst fabricating a small MKID array (which is produced using lithography) is a relatively straightforward task, ensuring significantly larger format arrays (target of Mpixel), potentially with a specific pixel layouts, requires investigation and development. 
\item Readout Electronics:
The current read-out electronics are for ground use and have not been optimised. The electronics, which are a significant part of the imaging system, will need to be re-evaluated, customised for large arrays and designed for space use.
\item Cooling system:
The MSSL continuous sub-Kelvin cryo-cooler is based on space qualified heritage \cite{BARTLETT2010582}
but is currently not designed for space. Analysis to identify modifications for flight is required initially, followed ultimately by qualification. A complete thermal analysis and mass/power trade-off is also needed to determine the most suitable pre-cooling chain,
whether that is liquid helium, cryo-coolers or a combination (e.g. Herschel, Planck, Astro-H).
\end{itemize}

\section{Conclusions}\label{conclusions}
In this white paper we have briefly outlined some technologies capable of achieving all-sky NIR Gaia-like astrometry. Such technology may have other applications in astronomy, remote sensing, planetary observation and for LIDAR technology. Four approaches have been identified each with their own unique challenges. 
\begin{itemize}[itemsep=0.1mm, parsep=0pt]
\item The hybrid solution using a HgCdTe layer bump bonded to a Si CCD must demonstrate that charge can be efficiently transferred to the silicon CCD layer and show that the two materials can be operated effectively at the same temperature in a space environment.
\item HgCdTe APDs with TDI signal processing technology are very promising and rapidly developing but its scaling to large format arrays needed for space astrometry must also be demonstrated. 
\item Ge detectors are very promising new devices for our application but must show that the technology can mature sufficiently quickly to meet our needs for large format detectors. 
\item MKIDs are very interesting with high time resolution, moderate resolving power, and natural multiplexing but scaling MKID detectors for space applications is a challenge and, 
in particular, moving towards Giga-pixel space based devices is very challenging given that they also require active cooling.
\end{itemize}
We have briefly reviewed the state of the profession for scanning NIR detectors. There are clear ways forward but this will require direct investment (Small: $<$ \$500M) by space agencies, funding agencies and private companies to enable this technology for space applications. The purpose of this white paper is to highlight this developing field to ensure this technology is available for space and other applications. Laboratory studies are needed in the short term before investment in full scale production of such devices can begin.

\newpage
\bibliographystyle{unsrt}
\bibliography{DetectorWhitePaper}

\begin{thebibliography}{10}

\bibitem{2016arXiv160907325H}
David {Hobbs}, Erik {H{\o}g}, Alcione {Mora}, Cian {Crowley}, Paul {McMillan},
  Piero {Ranalli}, Ulrike {Heiter}, Carme {Jordi}, Nigel {Hambly}, Ross
  {Church}, Brown {Anthony}, Paolo {Tanga}, Laurent {Chemin}, Jordi {Portail},
  Fran {Jim{\'e}nez-Esteban}, Sergei {Klioner}, Francois {Mignard}, Johan
  {Fynbo}, {\L}ukasz {Wyrzykowski}, Krzysztof {Rybicki}, Richard~I. {Anderson},
  Alberto {Cellino}, Claus {Fabricius}, Michael {Davidson}, and Lennart
  {Lindegren}.
\newblock {GaiaNIR: Combining optical and Near-Infra-Red (NIR) capabilities
  with Time-Delay-Integration (TDI) sensors for a future Gaia-like mission}.
\newblock {\em arXiv e-prints}, page arXiv:1609.07325, Sep 2016.

\bibitem{2019arXiv190408836M}
Barbara {McArthur}, David {Hobbs}, Erik {H{\o}g}, Valeri {Makarov}, Alessandro
  {Sozzetti}, Anthony {Brown}, Alberto {Krone Martins}, Jennifer~Lynn
  {Bartlett}, John {Tomsick}, Mike {Shao}, Fritz {Benedict}, Eduardo {Bendek},
  Celine {Boehm}, Charlie {Conroy}, Johan~Peter {Uldall Fynbo}, Oleg {Gnedin},
  Lynne {Hillenbrand }, Lennart {Lindegren}, David~R. {Rodriguez}, Rick
  {White}, Slava {Turyshev}, Stephen {Unwin}, and ChengXing {Zhai}.
\newblock {All-Sky Near Infrared Space Astrometry}.
\newblock {\em arXiv e-prints}, page arXiv:1904.08836, Apr 2019.

\bibitem{LeitzGe2018}
C.~W. {Leitz}, M.~{Zhu}, S.~{Rabe}, B.~{Burke}, I.~{Prigozhin}, D.~{O'Mara},
  K.~{Ryu}, M.~{Cooper}, R.~{Reich}, K.~{Johnson}, W.~{Hu}, B.~{Felton},
  M.~{Cook}, {Stull} C., and {Suntharalingam} V.
\newblock Development of germanium charge-coupled devices.
\newblock In {\em SPIE Digital Library}, volume 10709 of {\em Proceedings of
  SPIE}, 2018.

\bibitem{DOI:10.14429}
A.~{Singh} and R.~{Pal}.
\newblock Infrared avalanche photodiode detectors.
\newblock {\em Defence Science Journal}, 67(2):159--168, 2017.

\bibitem{Feautrier:15}
P.~{Feautrier} and J.~{Gach}.
\newblock State of the art ir cameras for wavefront sensing using e-apd mct
  arrays.
\newblock 2015.

\bibitem{Sun:17}
X.~{Sun}, J~B. {Abshire}, J.~D. {Beck}, P.~{Mitra}, K.~{Reiff}, and G.~{Yang}.
\newblock Hgcdte avalanche photodiode detectors for airborne and spaceborne
  lidar at infrared wavelengths.
\newblock {\em Opt. Express}, 25(14):16589--16602, Jul 2017.

\bibitem{Leonardo:16}
I.~{Baker}, C.~{Maxey}, L.~{Hipwood}, and K.~{Barnes}.
\newblock Leonardo (formerly selex es) infrared sensors for astronomy: present
  and future.
\newblock In {\em SPIE Digital Library}, volume 9915 of {\em Proceedings of
  SPIE}, 2016.

\bibitem{Atkinson:2014}
D.~E. {Atkinson}, D.~N.~B. {Hall}, C.~{Baranec}, I.~M. {Baker}, S.~M.
  {Jacobson}, and R.~{Riddle}.
\newblock Observatory deployment and characterization of saphira hgcdte apd
  arrays, 2014.

\bibitem{Atkinson:2016}
D.~E. {Atkinson}, D.~N.~B. {Hall}, I.~M. {Baker}, S.~B. {Goebel}, S.~M.
  {Jacobson}, C.~{Lockhart}, and E.~A. {Warmbier}.
\newblock Next-generation performance of saphira hgcdte apds.
\newblock In {\em SPIE Digital Library}, volume 9915 of {\em Proceedings of
  SPIE}, 2016.

\bibitem{2018AJ....155..220A}
D.~{Atkinson}, D.~{Hall}, S.~{Jacobson}, and I.~M. {Baker}.
\newblock {Photon-counting Properties of SAPHIRA APD Arrays}.
\newblock {\em Astrophysical Journal}, 155:220, May 2018.

\bibitem{Atkinson:2018}
D.~E. {Atkinson}, D.~N.~B. {Hall}, S.~B. {Goebel}, S.~M. {Jacobson}, and I.~M.
  {Baker}.
\newblock Observatory deployment and characterization of saphira hgcdte apd
  arrays, 2018.

\bibitem{2014SPIE.9148E..17F}
G.~{Finger}, I.~{Baker}, D.~{Alvarez}, D.~{Ives}, L.~{Mehrgan}, M.~{Meyer},
  J.~{Stegmeier}, and H.~J. {Weller}.
\newblock {SAPHIRA detector for infrared wavefront sensing}.
\newblock In {\em Adaptive Optics Systems IV}, volume 9148 of {\em Proceedings
  of SPIE}, page 914817, August 2014.

\bibitem{2016SPIE.9909E..12F}
G.~{Finger}, I.~{Baker}, D.~{Alvarez}, C.~{Dupuy}, D.~{Ives}, M.~{Meyer},
  L.~{Mehrgan}, J.~{Stegmeier}, and H.~J. {Weller}.
\newblock {Sub-electron read noise and millisecond full-frame readout with the
  near infrared eAPD array SAPHIRA}.
\newblock In {\em Adaptive Optics Systems V}, volume 9909 of {\em Proceedings
  of SPIE}, page 990912, July 2016.

\bibitem{2016SPIE.9915E..0OH}
D.~N.~B. {Hall}, I.~{Baker}, and G.~{Finger}.
\newblock {Towards the next generation of L-APD MOVPE HgCdTe arrays: beyond the
  SAPHIRA 320 x 256}.
\newblock In {\em High Energy, Optical, and Infrared Detectors for Astronomy
  VII}, volume 9915 of {\em Proceedings of SPIE}, page 99150O, July 2016.

\bibitem{2018SPIE10702E..1VS}
R.~{Sharp}, D.~{Adams}, G.~{Bloxham}, R.~{Boz}, D.~{Bundy}, D.~{Chandler},
  G.~{Gausachs}, L.~{Gers}, J.~{Hart}, N.~{Herrald}, J.~{Nielsen},
  E.~{O'Brien}, C.~{Onken}, I.~{Price}, A.~{Vaccarella}, C.~{Vest}, and
  P.~{Young}.
\newblock {Design evolution of the Giant Magellan Telescope Integral Field
  Spectrograph, GMTIFS}.
\newblock In {\em Ground-based and Airborne Instrumentation for Astronomy VII},
  volume 10702 of {\em Society of Photo-Optical Instrumentation Engineers
  (SPIE) Conference Series}, page 107021V, July 2018.

\bibitem{2017SPIE10209E..0GF}
P.~{Feautrier}, J.-L. {Gach}, T.~{Greffe}, F.~{Clop}, S.~{Lemarchand},
  T.~{Carmignani}, E.~{Stadler}, C.~{Doucour{\'e}}, and D.~{Boutolleau}.
\newblock {C-RED One and C-RED 2: SWIR advanced cameras using Saphira e-APD and
  Snake InGaAs detectors}.
\newblock In {\em Society of Photo-Optical Instrumentation Engineers (SPIE)
  Conference Series}, volume 10209 of {\em Society of Photo-Optical
  Instrumentation Engineers (SPIE) Conference Series}, page 102090G, April
  2017.

\bibitem{96593b97fd7e4d50bb6c941d1f28c48d}
Derek Kopon, Brian McLeod, {Marcos A.} {Van Dam}, Antonin Bouchez, Ken
  McCracken, Daniel Catropa, William Podgorski, Stuart McMuldroch, Alan Conder,
  {Laird M} Close, Jared Males, Katie Morzinski, and Timothy Norton.
\newblock On-sky demonstration of the gmt dispersed fringe phasing sensor
  prototype on the magellan telescope.
\newblock In {\em Adaptive Optics Systems V}, volume 9909, United States, 2016.
  SPIE.

\bibitem{2018SPIE10709E..23V}
A.~{Vaccarella}, R.~{Sharp}, M.~{Ellis}, A.~{Bouchez}, R.~{Conan}, R.~{Boz},
  D.~{Bundy}, G.~{Gausachs}, J.~{Gilbert}, L.~{Gers}, J.~{Hart}, N.~{Herrald},
  M.~{Ireland}, J.~{Nielsen}, I.~{Price}, C.~{Vest}, and H.~{Zovaro}.
\newblock {Cryogenic detector preamplifer developments at the ANU}.
\newblock In {\em High Energy, Optical, and Infrared Detectors for Astronomy
  VIII}, volume 10709 of {\em Society of Photo-Optical Instrumentation
  Engineers (SPIE) Conference Series}, page 1070923, July 2018.

\bibitem{El-Desouki:14}
Munir~M. {El-Desouki} and Badeea {Al-Azem}.
\newblock Novel cmos time-delay integration using single-photon counting for
  high-speed industrial and aerospace applications, 2014.

\bibitem{Li:19}
S.~{Li}, J.~{Maire}, M.~{Cosens}, and S.~A. {Wright}.
\newblock Detector characterization of a near-infrared discrete avalanche
  photodiode 5x5 array for astrophysical observations.
\newblock volume 11002, 2019.

\bibitem{2018SPIE10702E..5IW}
Shelley~A. {Wright}, Paul {Horowitz}, J{\'e}r{\^o}me {Maire}, Dan {Werthimer},
  Franklin {Antonio}, Michael {Aronson}, Sam {Chaim-Weismann}, Maren {Cosens},
  Frank~D. {Drake}, and Andrew~W. {Howard}.
\newblock {Panoramic optical and near-infrared SETI instrument: overall
  specifications and science program}.
\newblock In {\em Ground-based and Airborne Instrumentation for Astronomy VII},
  volume 10702 of {\em Society of Photo-Optical Instrumentation Engineers
  (SPIE) Conference Series}, page 107025I, Jul 2018.

\bibitem{1974ApPhL..25..747S}
D.~K. {Schroder}.
\newblock {A two-phase germanium charge-coupled device}.
\newblock {\em Applied Physics Letters}, 25:747, December 1974.

\bibitem{Leitz_2017}
C.~{Leitz}, S.~{Rabe}, I.~{Prigozhin}, B.~{Burke}, {Zhu} M., K.~{Ryu},
  M.~{Cooper}, R.~{Reich}, K.~{Johnson}, W.~L. {Hu}, B.~{Felton}, M.~{Cook},
  C.~{Stull}, and V.~{Suntharalingam}.
\newblock Germanium {CCDs} for large-format {SWIR} and x-ray imaging.
\newblock {\em Journal of Instrumentation}, 12(05):C05014--C05014, May 2017.

\bibitem{doi:10.1063/1.3292300}
B.~A. {Mazin}.
\newblock Microwave kinetic inductance detectors: The first decade.
\newblock {\em AIP Conference Proceedings}, 1185(1):135--142, 2009.

\bibitem{2017OExpr..2525894S}
P.~{Szypryt}, S.~R. {Meeker}, G.~{Coiffard}, N.~{Fruitwala}, B.~{Bumble},
  G.~{Ulbricht}, A.~B. {Walter}, M.~{Daal}, C.~{Bockstiegel}, G.~{Collura},
  N.~{Zobrist}, I.~{Lipartito}, and B.~A. {Mazin}.
\newblock {Large-format platinum silicide microwave kinetic inductance
  detectors for optical to near-IR astronomy}.
\newblock {\em Optics Express}, 25(21):25894, Oct 2017.

\bibitem{2012SPIE.8452E..1OB}
J.~{Bartlett}, G.~{Hardy}, I.~{Hepburn}, S.~{Milward}, P.~{Coker}, and
  C.~{Theobald}.
\newblock {Millikelvin cryocooler for space- and ground-based detector
  systems}.
\newblock In {\em Millimeter, Submillimeter, and Far-Infrared Detectors and
  Instrumentation for Astronomy VI}, volume 8452 of {\em Proceedings of SPIE},
  page 84521O, September 2012.

\bibitem{Mazin2019}
B.~{Mazin} and et~{al}.
\newblock Mkids in the 2020s.
\newblock {\em ASTRO-2020 APC White Paper}, 2019.

\bibitem{BARTLETT2010582}
J.~{Bartlett}, G.~{Hardy}, I.~D. {Hepburn}, C.~{Brockley-Blatt}, P.~{Coker},
  E.~{Crofts}, B.~{Winter}, S.~{Milward}, R.~{Stafford-Allen}, M.~{Brownhill},
  J.~{Reed}, M.~{Linder}, and N.~{Rando}.
\newblock Improved performance of an engineering model cryogen free double
  adiabatic demagnetization refrigerator.
\newblock {\em Cryogenics}, 50(9):582 -- 590, 2010.
\newblock 2009 Space Cryogenic Workshop.

\end{thebibliography}

\end{document}